# The upgraded data acquisition system for beam loss monitoring at the Fermilab Tevatron and Main Injector[*]

A. Baumbaugh, C. Briegel, B. C. Brown, D. Capista, C. Drennan, B. Fellenz,
K. Knickerbocker,[†] J. D. Lewis, A. Marchionni,[‡] C. Needles,[§] M. Olson,[**]
S. Pordes, Z. Shi, D. Still, R. Thurman-Keup, M. Utes and J. Wu

*Fermi National Accelerator Laboratory,
Box 500, Batavia, Illinois, 60510, USA*

   *E-mail*: molson@fnal.gov

ABSTRACT: A VME-based data acquisition system for beam-loss monitors has been developed and is in use in the Tevatron and Main Injector accelerators at the Fermilab complex. The need for enhanced beam-loss protection when the Tevatron is operating in collider-mode was the main driving force for the new design. Prior to the implementation of the present system, the beam-loss monitor system was disabled during collider operation and protection of the Tevatron magnets relied on the quench protection system. The new Beam-Loss Monitor system allows appropriate abort logic and thresholds to be set over the full set of collider operating conditions. The system also records a history of beam-loss data prior to a beam-abort event for post-abort analysis. Installation of the Main Injector system occurred in the fall of 2006 and the Tevatron system in the summer of 2007. Both systems were fully operation by the summer of 2008. In this paper we report on the overall system design, provide a description of its normal operation, and show a number of examples of its use in both the Main Injector and Tevatron.

---

[*] Work supported by Fermi Research Alliance, LLC under Contract No. DE-AC02-07CH11359 with the United States Department of Energy.
[†] Present Address: Los Alamos National Laboratory, Los Alamos, New Mexico, USA.
[‡] Present Address: ETH Institute for Particle Physics, ETH Zurich, Zurich, Switzerland.
[§] Retired.
[**] Corresponding author.

**Contents**

## 1. Introduction

A project to upgrade the Beam-Loss Monitor systems in the Fermilab Main Injector and the Tevatron was launched in 2004. The motivation in the case of the Main Injector was to develop a system capable of providing data to avoid excessive activation of components and thus allow the pursuit of higher intensity. The primary motivation at the Tevatron was to enable the beam-loss monitor system which had been designed for fixed-target operation to operate during collider operation. The Tevatron synchrotron at Fermilab is a superconducting proton and antiproton collider running at an energy of 980 GeV per beam. The stored energy in the beam is such that the inadvertent loss of the beam can not only quench the superconducting magnets, but physically damage them. In collider operation, prior to the upgrade, the magnets were protected by a Quench Protection Monitor (QPM) that senses the development of a quench by sampling the voltage across each magnet and shunts the current through a dump resistor.

    In 2003, a moveable beamline detector called a roman pot was unintentionally inserted into the beam during collider operation causing a massive quench and damage to some of the collimators [1] (see figure 1).

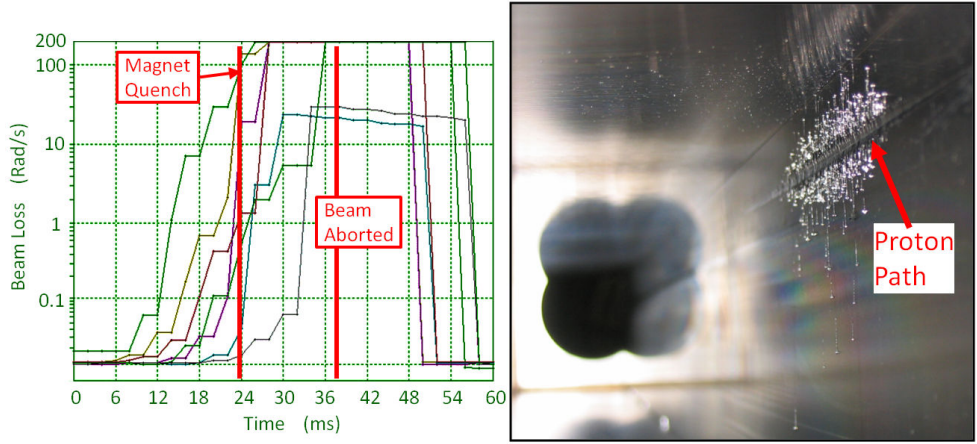

**Figure 1.** Damage to a collimator caused by the proton beam. Left) Plot of beam losses at a number of locations and the corresponding timing of a magnet quench and the eventual beam abort driven by the quench protection system. Right) Valley carved by the protons in a 1.5 m collimator system absorber. The vertical streaks of melted steel emanating from the proton path are clearly visible.

As can be seen in the figure, the loss monitors were registering losses well before the quench occurred and could have pulled the abort much sooner than the QPM which eventually aborted the beam after the quench. Unfortunately, the original BLM system [2-5] was not enabled in the abort system when the Tevatron was at maximum energy since it had a non-negligible probability of generating false aborts. False aborts are a problem for the Tevatron because it uses antiprotons as one of the colliding particles and antiproton production is a relatively slow process (12 hours or more). Aborting the beam, therefore, means waiting a significant amount of time for more antiprotons to be produced. It was felt that, given the time it could take for a serious loss to develop, the QPM could protect the magnets adequately while avoiding any false aborts from the BLM system.

This quench incident however was not a normal quench. It was the first time a "fast quench" had happened where the QPM was unable to respond quickly enough. This led to improvements in the QPM and drove the effort for a replacement for the BLM system that was more flexible and less prone to accidental aborts.

The new BLM system is designed with several key features. It provides a flexible and reliable abort system to protect Tevatron magnets. Whereas the original BLM system had two abort configurations, the new system can support many that are switchable via machine state messages. The new system also supports multiple integration periods each of which can generate an abort. The upgrade provides loss monitor data during normal operations without interruption of the abort protection, and it provides detailed diagnostic loss histories when an abort happens.

The Main Injector also benefits from this new loss monitor system. As the beam intensities get higher, the need to control losses and the associated radioactivation of beamline elements becomes more critical. In addition to better measurement capabilities allowing for more detailed loss measurements, the abort functionality serves as a way to enforce the minimization of losses.

## 2. System overview

The accelerator complex at Fermilab consists of eight accelerators used for the purpose of performing particle physics experiments [6]. The beam loss monitor systems for the two high-energy accelerators, the Tevatron and Main Injector, use a set of ionization chambers inside the tunnel with the high voltage supply and all the signal processing electronics located outside the tunnel in service buildings. These systems were upgraded between 2006 and 2008. The upgrade involved replacing the HV supplies and all the signal processing electronics; the loss monitors themselves were not changed.

### 2.1 Loss monitor

The physical particle detector is unchanged from the original BLM system and the description here is paraphrased from [3]. The detector is an Ar-filled glass cylindrical ionization chamber with nickel electrodes and was chosen to be extremely radiation hard. The gas is 1 atm pure Ar with an active volume about 110 cm$^3$. No $CO_2$ was used, because $CO_2$ could eventually dissociate under ionizing radiation and alter the response. The detector calibration is ~70 nC/Rad and is extremely stable.

The ionization chamber design was a compromise among high gain, fast response, size, and low cost. The design included placing the anode (center electrode, positive high voltage) and cathode (outer electrode, signal) feedthroughs at opposite ends of the glass bottle with a guard ring on the glass envelope to minimize the dark current leakage (roughly 100 pA at 2500 V). Pulsed-beam measurements at Argonne National Lab [7] showed that the chamber could detect an instantaneous radiation dose (in 1 or 2 microseconds) of about 10 Rads with less than 20% charge loss. This large amount of space charge (about 700 nC) severely modifies the electric fields in the gas as the ions and electrons drift toward the electrodes, leading to both recombination losses and gas multiplication. For comparison, the inter-electrode capacitance is about 2 pF, leading to a displacement charge on the electrodes of about 5 nC at 2500 V.

The signal from the cathode is brought out of the tunnel enclosure on RG-58 cable, which is not particularly noise resistant, but is part of the original BLM system and was not replaced. Complicating the noise situation is the legacy high voltage distribution which is supplied via a single RG-58 cable daisy chained through typically 12 ionization chambers producing myriad ground loops with the signal cables. Additionally, noise sources vary depending on the location within the accelerator.

To combat the noise conditions, a variety of schemes are utilized to reduce the noise pickup to acceptable levels. An inductive choke is used on each signal cable just before the acquisition electronics to reduce large common-mode noise pickup on the cable. The size of the choke was chosen to give the proper frequency response. In the signal cables of the Main Injector systems, the ground connection is broken by a 100 kΩ resistor at the ionization chamber. This theoretically prevents ground loops and, in fact, greatly reduces the noise in the Main Injector. On the other hand, breaking the ground in the Tevatron has no effect; a fact which has never been understood other than to invoke a different noise source. In parallel with the hardware efforts at noise reduction, a software algorithm was developed to subtract coherent noise from the signal in cases where the analog noise rejection techniques were insufficient. This algorithm is discussed in section 3.1.

## 2.2 Data acquisition system

The new BLM system consists of a VME crate with a number of boards handling the various functions. Besides the VME Slot 1 crate processor card (CP), the BLM system hardware consists of a control card (CC) for setting up and controlling the overall state of the system, a timing card (TC) for generating and distributing timing signals to the other cards, an abort card (AC) containing the abort decision logic, a high voltage card (HV), and up to 15 four-channel integrator/digitizer cards (DC), each of which contains a circular buffer of loss readings and four sliding sums of varying length called Immediate, Fast, Slow, and Very Slow. All of these cards are individually described in the following sections.

The BLM system, as portrayed in figure 2, uses a standard 6Ux160mm VME format and communicates with the outside world via the Accelerator Controls Network (ACNET) through the CP.

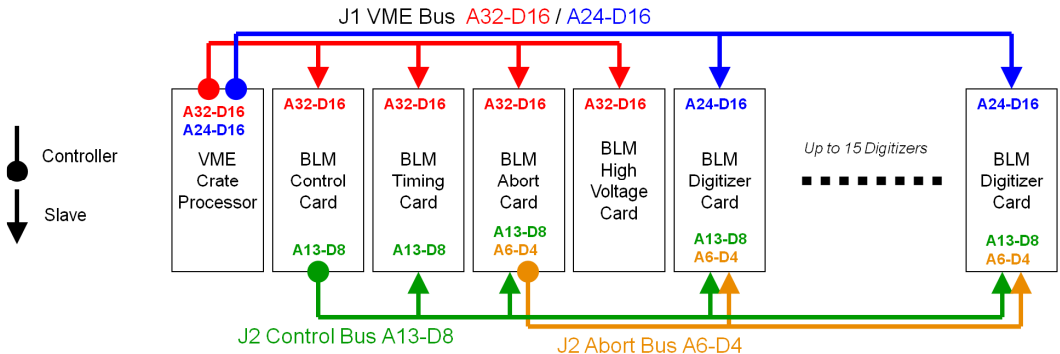

**Figure 2.** Block diagram of a BLM VME crate. Arrows indicate bus slave cards and circles indicate bus masters.

The VME crate has a standard J1 backplane and a custom J2 backplane. The J2 Backplane has a proprietary control bus as well as a proprietary abort bus in addition to VME 32-bit extensions. The CC is the master of the control bus. This bus has 13 address lines and eight data lines and handles all of the critical BLM controls. The control bus signals include:

- *Reset* - This signal is the BLM Crate reset. Driven by the CC, it resets the state machine on all the cards.
- *Reset_DC* - This signal clears the sliding sums and resets the circular buffer pointers. This signal typically is used to initialize the DC, TC, and AC and get ready for data taking.
- *Make_Meas* - The primary clock for the BLM system. It is generated on the TC and typically runs at a 21 μs period. It is derived either from the first bunch marker on the beam sync clock (see section 3.2) or from an internal clock in the event that the beam clock is not present. It is transmitted on the BLM control bus to all BLM cards. The shortest allowable period for this signal is 15 μs due to the reset time needed by the DC integrators. This signal causes the DC to switch integrators, latch the previous abort state, digitize the reading, reset the integrator, calculate the sliding sums, and make the abort comparisons. It also triggers the AC to check the abort status of each DC channel.
- *Fast_Latch* - This is a signal produced on the TC that tells the DC state machine to latch the current Fast sum into a register so the CC can read it.

- *Slow_Latch* - This is a signal produced on the TC that tells the DC state machine to latch the current Slow sum into a register so the CC can read it.
- *VSlow_Latch* - This is a signal produced on the TC that tells the state machine to latch the current Very Slow sum into a register so the CC can read it.

The AC is the master of the abort bus and the DCs are slaves. The abort bus consists of six address lines and four data lines. The abort bus functions as a way for the AC to poll the DC about abort conditions.
- *Abort_CS(0:5)* - Abort Channel select. These signals come from the AC and are used to poll all the DC channels to request abort conditions.
- *Abort(0:3)* - Abort lines 0,1,2,3. These are driven by the DC in response to *Abort_CS* and provide the AC the information about which, if any, of the four thresholds (Immediate, Fast, Slow, Very Slow) the requested channel is above.

The BLM system abort protection is handled entirely in FPGA code. The FPGA on the DC does the summing and comparisons, and the FPGA on the AC does the abort generation, and the timing of this is handled by the FPGA on the TC. This use of only FPGA code reduces the likelihood that a CPU lock up or crash will prevent the abort protection from functioning.

## 3. Hardware

### 3.1 Digitizer card

The digitizer card (DC) integrates and digitizes the current from up to four ionization chambers in ~21 μs periods. The sample period of 21 μs was chosen as it is the revolution period of the Tevatron and is also approximately twice the revolution period of the Main Injector. There is one TI/Burr-Brown ACF2101 dual-integrator chip per loss monitor running in a "ping-pong" mode [8]. One integrator is digitized and reset while the other is integrating leaving no gap in the collection of loss data. Once the integrated loss is digitized, the data are fed to an FPGA and used to construct the four sliding sums of user-defined time scales ranging from a single reading up to 1.4 seconds. These constructed data are compared against thresholds to generate the four abort signals.

The digitizer has 16-bit resolution, and the scaling is such that one digitizer count represents 15.26 fC of charge in the integrator. This corresponds to 0.22 μRad given the calibration of 70 nC/Rad. This dynamic range was chosen to cover a loss range from 0.01 to 10 times the quench value in the Tevatron. The actual dynamic range is $> 10^4$ to accommodate uncertainties in converting from the actual loss in the magnet to the loss seen by the ionization chamber.

To compensate for both intentional and unintentional offsets, a pedestal value is measured prior to each beam injection. It is digitized, and then subtracted from each integration period. In order to minimize the load on the CC, these subtractions are done by the CP.

The DC logic maintains four sliding sums per channel with programmable durations of up to 65,536 base clocks (1.4 s at 21 μs) The sliding sums are made by adding the present reading to the sum and subtracting the oldest reading from the sum. The sums are 32-bit integers readable though the control bus in four 8-bit segments. Since the Main Injector is a short cycle-time synchrotron (a few seconds), one of the sums can be set to an infinite period giving the total loss for the cycle. This "integration mode" requires special handling to avoid the possibility of underflows resulting in false aborts. It utilizes an optional squelch level to avoid integrating any possibly mis-measured pedestals; it performs pedestal subtraction; and it stores the values in 64-bit registers to avoid overflow from the possibly longer summation times.

The block diagram in figure 3 illustrates the signal processing for each channel [9]. Note that the sum registers are read and the threshold registers written over the control bus. The SRAM memory which stores the integrator output values can be read over the VME bus (J1) by the CP when the system is stopped.

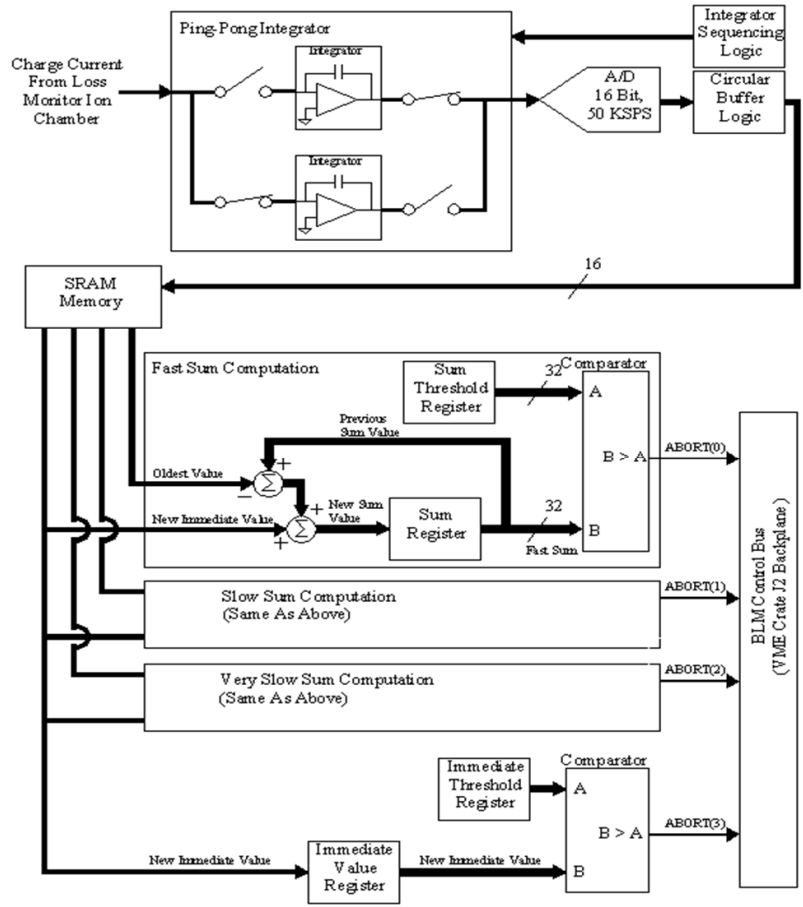

**Figure 3.** Block diagram of the Digitizer Card.

The Circular Buffer Logic block in figure 3 represents a DSP that performs significant manipulations of the digitized data to correct for noise present on the analog input signal [10]. The signal cables pick up 60Hz, 180Hz and multiples of 360Hz noise from equipment powered by three-phase 60Hz AC power as seen in figure 4. Calculating sliding sums can be viewed as a digital filtering process. The fast sliding sum (FS) with a 128 sample length shown in figure 5 has reduced the noise level from about 10 ADC counts in raw data down to about 2 ADC counts. Beam loss is apparent in the fast sliding sum trace.

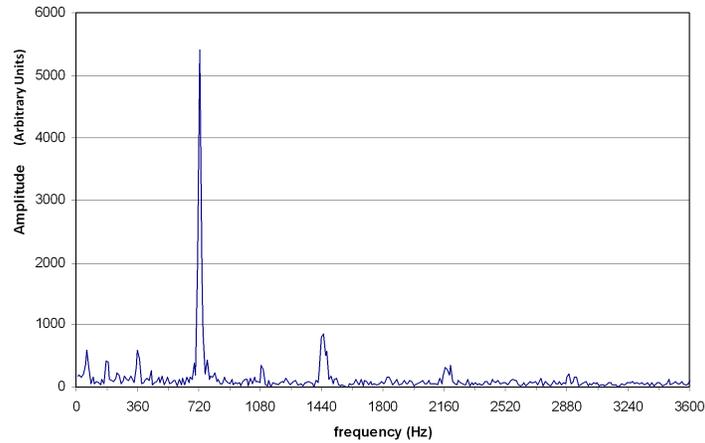

**Figure 4.** Typical noise spectrum observed in the Main Injector. One can clearly see the multiples of 60 Hz.

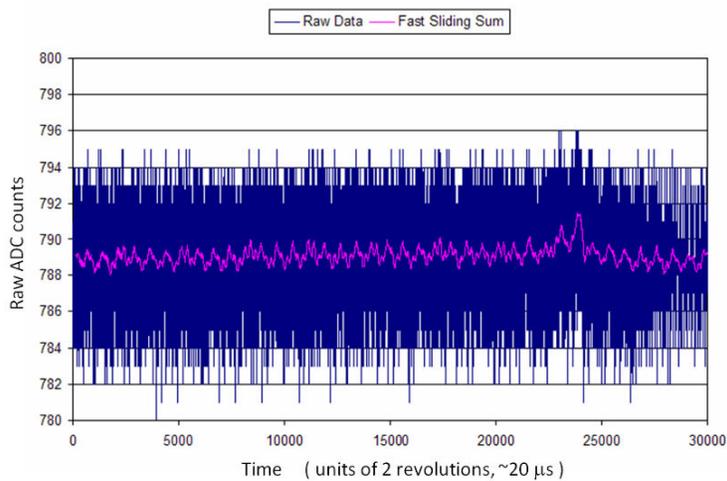

**Figure 5.** Example of raw measurement and fast sliding sum data from the Main Injector. The sum, which has been divided by the number of measurements, reduces the noise component as expected, revealing some of the underlying loss structure. However, the 60 Hz noise is still visible. This data has not had pedestals subtracted.

In places where the remaining noise in the sum is unacceptable, a de-ripple process using a Cascaded Integration Comb filter (CIC) can be employed to further reduce the noise component. A ripple waveform (WF) is extracted from the CIC filter and then subtracted from the signal to give the de-rippled loss signal (DR) as seen in figure 6. The waveform (WF) in general takes 3 periods to become valid. The de-rippled output first follows the CIC sum while WF is invalid. It then becomes a smoothed curve with harmonics of 60 Hz ripple canceled. Low level beam loss in now exposed.

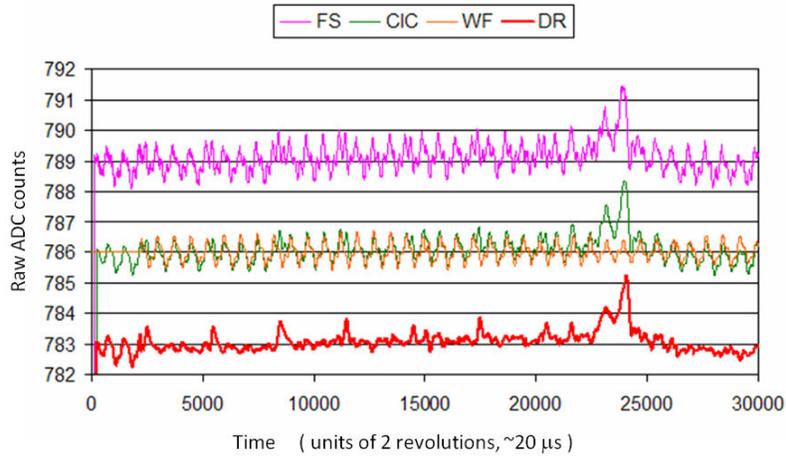

**Figure 6.** The De-ripple Process. The windowed sum (FS) has a significant amount of coherent noise present. The CIC filter is constructed to remove higher harmonics of 60 Hz. The noise waveform (WF) is constructed from zero-loss regions of the CIC signal and is subtracted from the signal leaving the desired de-rippled signal (DR) and revealing even more losses which were not obviously visible in the FS plot. (In order to see the traces clearly, an artificial offset has been added to each trace)

In cases of extremely large losses, the digitizer has a protection circuit to prevent the integrator from becoming saturated. During the integration phase of the cycle, the output voltage is checked by a comparator. If it reaches approximately 95% of full scale for the ADC, the input current is shut off, and the remaining charge is stored on the cable and is integrated on subsequent measurement periods. Thus for large losses, the leading edge time is preserved; however, the raw measurement will be near full scale until all charge from the loss has been collected. In this way, by summing over measurements, it is possible to record the full beam loss.

### 3.1.1 Calculations

This section presents the formulas for the various quantities calculated by the DC [11,12]. It also provides conversions from physical values to digitizer values which are needed for determining abort thresholds. In table 1, the highlighted quantities are calculations done by the DC. Other quantities are calculated by the CP or other entity. These other quantities could have been calculated by the DC but are not needed for the abort protection were not included in the interest of limiting the complexity of the FPGA code.

**Table 1.** Definitions of variables used in the equations describing the pedestals, sums, and filters employed in the BLM system. Highlighted quantities are calculated by the DC.

| Quantity | Description |
|---|---|
| $N_S$ | Number of integration cycles to skip before summing; this allows for stabilization of the integrator and is set to 256 as determined experimentally |
| $L_{PED}$ | Pedestal sum length; typically 795, which is one 60 Hz cycle, to avoid offsets from power line noise pickup |
| $T$ | Type of sum (F=Fast, S=Slow, VS=Very Slow) |
| $L_T$ | # of summed integration cycles for sum type $T$; typically 64 for fast, ~1600 for slow, and 50 000 for very slow. These numbers do not originate from any particular requirement; they are simply spaced somewhat logarithmically over the available range. |
| $L_C$ | CIC filter length; used in the de-rippled sum; typically ~130 which produces zeros in the frequency domain starting at 360 Hz |
| $X_i$ | Raw data measurement |
| $S_i^T$ | Sliding sum type $T$ |
| $S_i^D$ | De-rippled sum |
| $S_j^C$ | CIC sliding sum which forms part of the de-rippled sum |
| $P$ | Pedestal |
| $W_{j-j0}$ | Baseline waveform for subtraction in de-rippled output |
| $q$ | Squelch level in units of # of sigma above noise rms; the squelch has never been used in practice |
| $\sigma$ | Raw data measurement rms |
| $Q$ | Squelch level in units of 16 * the very slow sum |
| $Y_i$ | Integrated value on Digitizer Card (64 bits) |
| $Z_i$ | Integrated value in Control card circular buffer |
| $A_i^T$ | Average pedestal subtracted measurement using the specified sum in units of mRad/s |
| $A_i^C$ | Pedestal subtracted, de-rippled measurement in units of mRad/s |
| $I_i$ | Real integrated value in units of mRad |
| $P_I$ | Integration pedestal |
| $I_{Thresh}$ | Abort threshold for integrations in ADC counts |
| $A^T_{Thresh}$ | Abort threshold for sum type $T$ in ADC counts |
| $I_{Lim}$ | Abort threshold for integrations in mRad |
| $A^T_{Lim}$ | Abort threshold for sum type $T$ in mRad/s |

The pedestal starts after a delay of NS integration cycles to give the analog portion of the board time to settle. It is given by

$$P = \sum_{i=16N_S}^{(16N_S+L_{PED}-1)} X_i \ . \tag{3.1}$$

The de-rippled baseline waveform is collected from CIC-filtered data when there are no losses:

$$W_{j-j_0} = \sum_{k=(j-j_0)-L_C+1}^{j-j_0} \left( \sum_{i=k-L_C+1}^{k} X_i \right). \qquad (3.2)$$

The sliding sum type $T$ is the sum of the previous $L_T$ raw data measurements:

$$S_j^T = \sum_{i=j-L_T+1}^{j} X_i. \qquad (3.3)$$

The De-Rippled sum, $S_j^D$, is a sum of sums (CIC filter),

$$S_j^C = \sum_{k=j-L_C+1}^{j} \left( \sum_{i=k-L_C+1}^{k} X_i \right), \qquad (3.4)$$

minus the baseline waveform plus the average baseline,

$$S_j^D = S_j^C - W_{j-j_0} + \overline{W}. \qquad (3.5)$$

The squelch level that must be downloaded to the digitizer is ~16 times the very slow sum noise level times the number of sigma desired (it is 16*17 rather than 16*16 because of the contribution of the pedestal)

$$Q = q\sigma \sqrt{16 \times 17 \times L_{VS}} \qquad (3.6)$$

where the squelch level $q\sigma$ corresponding to a loss level, $R$ in mRad, can be obtained from

$$q\sigma = \frac{R\sqrt{L_F}}{10.4}. \qquad (3.7)$$

The integration mode integral starts accumulating after the delay and the pedestal measurement, and it updates every integration cycle. The integral has an offset of $2^{27}$ added to it to prevent negative numbers on the digitizer. The $U(\ )$ function is the unit step function ( $U(x) = 0$ if $x < 0$ and $= 1$ if $x > 0$ ):

$$Y_k = 2^{27} + \sum_{j=(16N_S+L_{PED})}^{k} (16 S_j^{VS} - P) \times U(16 S_j^{VS} - P - Q). \qquad (3.8)$$

The integration mode value is stored in the DC as a 64-bit number. The CC reads out bits 16-47 for a 32-bit loss value of

$$Z_k = \frac{Y_k}{2^{16}}. \qquad (3.9)$$

In the front end, the sums are turned into averages and have pedestals subtracted and conversion factors applied. For the Fast, Slow, and Very Slow type sums,

$$A_i^T = \left[ \frac{S_i^T}{L_T} - \frac{P}{L_{PED}} \right] \times 10.4 \quad \text{mRad/s} \qquad (3.10)$$

whereas for the De-Rippled,

$$A_i^C = \left[ \frac{S_i^C}{L_C^2} - \frac{P}{L_{PED}} \right] \times 10.4 \quad \text{mRad/s}. \qquad (3.11)$$

The integration mode value must be adjusted for the 16 bit offset, the factor of 16, and the over counting by $L_{VS}$ (see equations 3.8 and 3.9). A pedestal, $P_I = 2^{27}/2^{16}$, must also be subtracted:

$$I_i = \frac{(Z_i - P_I) \times 2^{16}}{16 L_{VS}} \times 0.000229 \quad \text{mRad} . \tag{3.12}$$

The abort thresholds are the inverse of the $A^T$ and $I$ values:

$$I_{Thresh} = \frac{I_{Lim} \times 16 L_{VS}}{2^{16} \times 0.000229 \text{ mRad}} + P_I \tag{3.13}$$

$$A^T_{Thresh} = \left[ \frac{A^T_{Lim}}{10.4 \text{ mRad/s}} + \frac{P}{L_{PED}} \right] \times L_T . \tag{3.14}$$

The abort thresholds in the Tevatron are set near the quench limit for energy deposition in the superconducting magnets which is somewhere around 50 Rads for an instantaneous deposit and 800 Rads/s for a slow deposition (derived from [3]). In the Main Injector, the abort thresholds are chosen to allow for some amount of tuning of the beam but low enough to prevent excessive losses from repeatedly occurring. However, in some cycles such as antiproton transfers to the Tevatron, the aborts are disabled to prevent the accidental loss of the antiprotons.

## 3.2 Timing card

The timing card (TC) receives accelerator timing information from three sources:
- Tevatron Clock (TCLK) – A 10 MHz clock signal that employs a modified Manchester code to generate up to 256 different clock events. These events are used for timing and synchronization of accelerator equipment.
- Beam Sync Clock (BSYNC) – A clock derived from an rf bucket counter. It is used to synchronize beam transfers into selected buckets. Every beam revolution it generates a marker pulse that is synchronized with rf bucket number 1.
- Machine Data (MDAT) – A 10 MHz clock with machine data such as bus current or beam intensity encrypted upon it. It uses 28-bit frames that represent eight bits of frame type, 16 bits of data and 4 bits of administrative overhead.

The TC decodes BSYNC to generate the 21 μs BLM system master clock which it distributes on the BLM control bus as the *Make_Meas* signal. The TC maintains a 64k circular buffer of timing information for each cycle including a 32-bit time of day and a 24-bit microsecond counter (reset at one second intervals); this buffer is synchronized with the circular buffer of loss measurements in the DCs. The master clock defines the integration interval of the DCs and sets the threshold comparison timing and abort logic comparison timing. It also generates the signals at the appropriate intervals that cause the DCs to latch the current values of the sliding sums and the CC to read and store these sums to the associated circular buffer.

The TC decodes TCLK and sends a signal to freeze the data buffers in the CC, TC, AC and DCs in the case of an abort. Other events from TCLK are used to signal the BLM system to collect and store synchronous ring-wide data samples for beam studies.

The TC uses the MDAT signal to determine the machine state and generate an interrupt to the CC so it can load the appropriate abort thresholds and logic when the accelerator state changes.

### 3.3 Control card

The control card (CC) communicates with the other system cards over the control bus, keeping local communications separate from VME data transfers. To ensure that data communications and other tasks running on the VME CP do not interrupt the continuous operation of the BLM abort logic, the control card provides a dedicated eZ80 processor that manages the setting of abort thresholds and other parameters used in the abort logic [12]. Communication between the CC and the CP is accomplished via dual-ported memory on the CC which is readable/writeable by both cards. The CC stores separate abort thresholds for all four sliding sums of every BLM channel for up to 256 different machine states. When a change in accelerator state is detected, the CC updates the thresholds in the DCs as well as the abort masks and multiplicity requirements in the AC.

The control card maintains circular buffers that store the histories of the Fast, Slow, and Very Slow sums for each digitizer channel with time stamps provided by the Timing Card. These histories are 16384 time bins deep for the Fast sums and 4096 deep for the Slow and Very Slow sums and are read out via VME on command from the CP.

The control card responds to the following 5 types of signals:
- TCLK Events - The CC polls the TCLK FIFO status register of the TC to determine if there is data in the FIFO. If there is data, it reads it from the FIFO and responds appropriately.
- MDAT - The CC polls the Timing Card's MDAT FIFO status register. If new data is there, it reads it, switches each DC's thresholds to the appropriate ones, copies the abort mask information corresponding to the MDAT machine state from VME memory to the AC, and tells the TC to generate an update abort settings signal.
- Data Latch Interrupt - At the end of each sum period, a corresponding latch is generated by the TC, interrupting the CC which reads the latch status registers in the TC to determine which sum data to read from the DC. The CC then reads the data and stores it in the circular buffers in VME accessible memory. Finally the CC clears the corresponding status register which in turn clears the interrupt.
- Abort_Service - The CC periodically polls the AC to determine the state of the AC and modifies the CC status register to reflect that state. If the AC is indicating an abort was requested, the snapshot of the last frame of the abort card before the abort is copied to VME memory for access by the CP. The cumulative OR of the snapshots is also copied to VME memory.
- Crate Processor - When settings need to be updated, the CP writes the settings to the CC and sets an appropriate register in the CC which then responds by loading the settings into the appropriate cards at the appropriate time.

### 3.4 Abort card

The abort card (AC), as depicted in figure 7, is master of the abort bus, which is used to gather abort data from the DCs in the crate [13]. Upon receiving a *Make_Meas* signal, the AC sends addresses out on the abort bus. For each address sent, it receives from a DC one BLM channel's four bits of abort data on the abort bus data lines. These four bits correspond to the four abort types: Immediate Abort (single turn loss over threshold), Fast Abort (Fast sum over threshold), Slow Abort (Slow sum over threshold), and Very Slow Abort (Very Slow sum over threshold). Upon receiving all the abort information, the AC adds the number of channel hits per abort type and compares the value to a preset multiplicity threshold. If a value is greater than the multiplicity threshold, the AC will send an abort request to the accelerator abort system fan-in box via a coaxial connector on the front panel. Each of the four abort types behaves independently in this manner. Each digitizer channel in the crate can be masked to prevent it from being included in an abort request decision. There are four masks per channel corresponding to the four abort types.

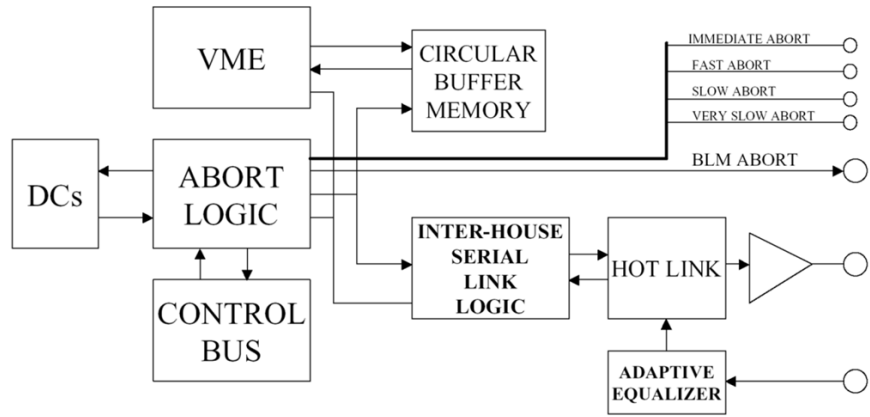

**Figure 7.** The block diagram of the abort card.

The AC uses control bus information from the CC to set the abort mask bits and to set the six-bit multiplicity value for each of the four abort types. To accommodate the different operating conditions, such as 150 GeV injection or 980 GeV colliding beam in the Tevatron, the abort masks and multiplicity thresholds in the AC are changed depending on the machine state. An important feature of the BLM system is that all abort operations are handled by state machines. Once setup by the CC these operations proceed without intervention from either VME or the CC.

A secondary BLM abort condition unique to the Tevatron uses a serial link between service buildings to monitor the abort status of all loss monitors in the Tevatron. This ring-wide loss analysis capability was added after observation of a quench revealed that prior to the quench, a number of BLMs at scattered aperture restrictions around the ring showed low-level losses. This serial link allows an abort multiplicity of loss monitors that are not in the same service building. The link uses a heliax cable run between buildings to transmit the data. A specially configured AC in one BLM crate serves as the collection point and processing center of the information. This card sends a "preamble" to an adjacent building which recognizes this and puts its 256 bits of abort status onto the data stream. Each crate receives the data stream from the previous one, appends its information to the end of the data stream, and transmits the appended stream to the next service building. When the data stream makes it all the way around the ring, there is a preamble and 27 crates worth of data which is analyzed and the abort is pulled if appropriate. The serial link utilizes a Cypress Hot Link transceiver operating at 60 Mb/s. Because of the cable lengths involved, the abort card has an equalizer circuit to compensate for the degradation of high-frequency components due to the skin effect. At 60 Mb/s, the 9600 bits require 160 µs to traverse the entire ring.

The VME section of the AC provides control/status for diagnostic control, read of the abort history circular buffer, circular buffer memory test, and inter-service building serial link tests.

### 3.5 High voltage card

The high voltage card provides the high voltage bias for the loss-monitor chambers. The high voltage is settable via a VME command and is normally set to 2000 V. The high voltage card is not a member of the J2 control bus. It has on board:

- An FPGA-based VME interface through which the CP can read and set voltages and currents among other things.
- A 16:1 multiplexer connected to a 16-bit ADC for digitization of all the monitored signals.
- Up to four high-voltage power supplies (DC-DC converters) each controlled by a 0-10 V programmable voltage supplied by a 12-bit digital to analog converter. The high-voltage supply is an Emco G30, which supplies 0 to 3000 V with a maximum current of 500 μA, < 200 mV$_{pp}$ ripple and better than 0.1% regulation at full load. The high-voltage supply is enclosed in an aluminum module with two SHV connectors: one for high voltage output and one for high voltage return. A third connector is used for I/O interface.

### 4. Normal operation

In order to smoothly update the new settings in the DCs and AC, these cards must double buffer all registers with the first register being written via the BLM control bus. The data is transferred to the actual usage register via a backplane update signal which occurs after all settings have been written and is synchronous with the *Make_Meas* signal. This update signal is generated on the TC so as to be synchronized with the *Make_Meas* and is generated in response to a command from the CC.

Once the CC loads the settings into all the cards on the control bus, the system is ready to run. The BLM operations are initiated by a prepare-for-beam clock event which causes the CC to tell the TC to issue a *Reset_DC* on the control bus. The *Reset_DC* causes the DCs to zero all sliding sums and causes the DCs, TC, and AC to reset all circular buffer pointers. This assures that all buffers are synchronized and ready to take data.

The TC, in addition to being the source of the *Make_Meas* signal, stores real-time clock data on each *Make_Meas* in a 64k deep circular buffer that is synchronized with those of the DCs and AC.

On the DCs, the *Make_Meas* signal defines the sample period, causing the integrators to toggle modes and triggering the ADCs to digitize the charge of the integrator not currently integrating. After that, the sliding sums are updated and all abort comparisons are made. At this time the new ADC readings are written to a 64k deep circular buffer which is used for diagnostic purposes as well as the source of the sliding sums. The new ADC data may also be written to one of two turn-by-turn (TBT) dedicated studies buffers. The abort states are latched on the next *Make_Meas*. Thus the DC has the full sample period to do its conversions, make the sliding sums, and do the abort comparison with thresholds.

On the AC, the *Make_Meas* signal causes the abort summing state machine to cycle through each BLM channel by putting the channel address on the abort bus and reading back from each channel the state of the threshold comparison. For each abort type, each channel has an abort mask bit, set by the CC, which determines if that channel is allowed to request an abort of that type. If the number of allowed over-threshold channels for a given abort type exceeds the abort multiplicity setting for that abort type, an abort request is transmitted from the card on a 50Ω TTL line driver. Note that while a sliding sum might be a sum over 500 samples (10ms) its abort threshold is compared every 21 µs.

During each 21 µs cycle, the DCs update the three sliding sums. For diagnostic purposes, these sums are stored periodically in circular buffers on the control card. This process is controlled by the TC, which periodically generates three latch signals, one for each sliding sum. The latch signals cause the DCs to latch the appropriate sum and the TC to latch the time stamp and to interrupt the CC so that it knows the data is latched and ready to be read and stored in the appropriate circular buffer. The individual ADC readings are 16-bits; however, the sliding sums are 32-bit numbers. Therefore, the dynamic range of, for example, a 1 second sliding sum is almost 32 bits or almost 900 Rads. These sliding sums are the total integrated loss over the sum interval, not just samples of losses spaced in time.

In the Tevatron, at any given time, the BLM has a variety of stored loss histories with different time resolutions: the 64k deep raw measurement buffer provides 1.4 seconds of loss data with 21µs resolution; the 16k Fast circular buffer provides 16 seconds of integrated loss data with 1ms resolution, the 4k Slow circular buffer provides 200 seconds of integrated loss data with 50ms resolution; and the 4k Very Slow buffer provides 4096 seconds, over an hour, of integrated loss data with 1 second resolution. In the event of an abort, there is a very detailed history of losses prior to the abort, which may be examined to aid in diagnosing the problem.

In the Main Injector, the total loss for each cycle is logged offline and used in diagnosing long term changes in the behavior of the Main Injector.

## 5. Operational Performance

### 5.1 Tevatron

The upgraded Tevatron BLM system became fully operational July 27, 2008. It incorporated all 289 ring BLM's with the 27 new VME BLM crates. The system provides BLM data for the Tevatron which can be acquired and plotted for all or individual BLMs. It is configured with abort level thresholds to provide protection from excessive loss, and in the event a BLM loss threshold was exceeded, the beam would be dumped or aborted by a set of single-turn kicker magnets.

Figure 8 depicts the loss pattern during an intentional abort of the Tevatron beam. The data is taken from the fast beam loss snapshot buffer so each frame is a ~1 ms interval. The proton and antiproton abort beam absorbers are located as shown in the figure. As the abort kickers are fired and the beams are absorbed in the dumps, the loss monitors in the vicinity of the dumps register large losses. Figure 9 depicts the turn-by-turn losses on a single loss monitor during the same beam abort as figure 8.

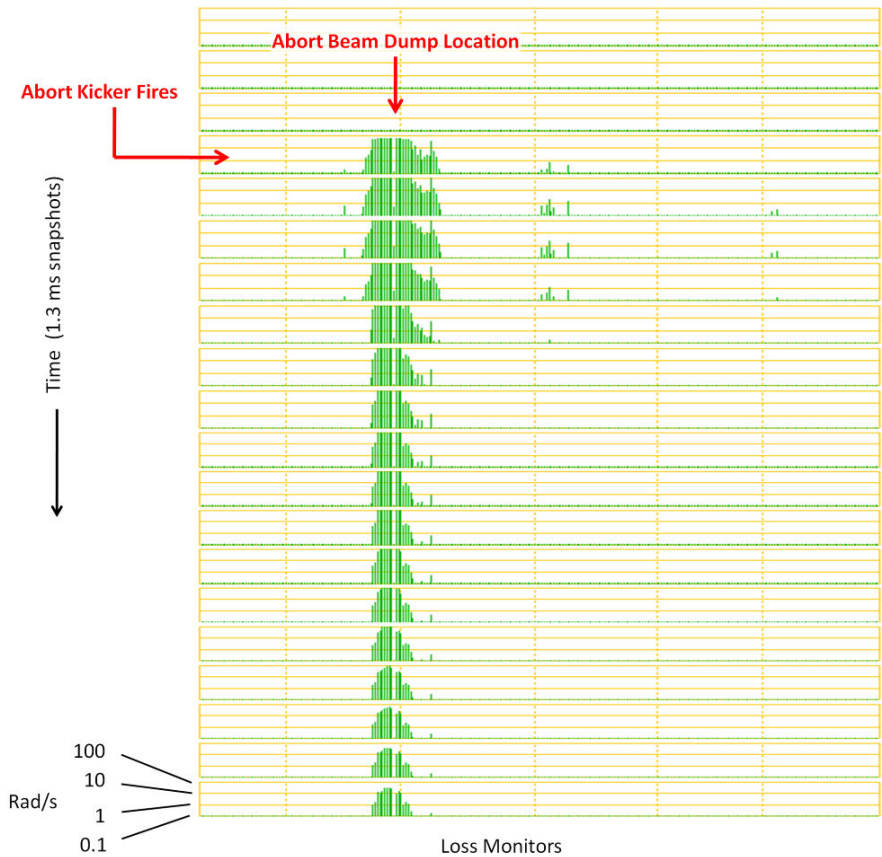

**Figure 8.** Tevatron aborted beam snapshot frames. Time runs from top to bottom with each successive logarithmic plot containing 1.3 ms of integrated losses. Each histogram bar is an individual loss monitor. The protons travel to the right on this plot while the antiprotons travel to the left.

The abort kickers remove the beam from the Tevatron in a single turn, so all the losses from aborting the beam occur in that one turn. From figure 8, one can see that the time to integrate all the losses is many milliseconds, which is much longer than a single 21 μs revolution. One can also see from figure 9 that while the total integration is slow, the response happens very quickly, nominally within one revolution. This allows for large single turn losses to generate aborts.

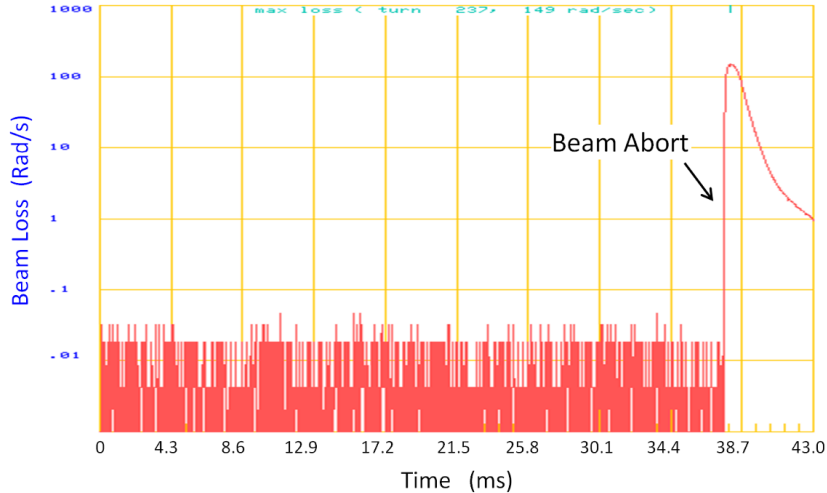

**Figure 9.** Turn-by-turn data of a single loss monitor for a normal beam abort. The rising edge of the losses is very fast and allows for a single-turn loss abort.

Initially, the BLM system was configured to allow aborts at 150 GeV injection and 150 GeV circulating beam conditions. Operationally in the Tevatron, only the Slow and Fast modes are configured while the Very Slow and Immediate integration modes are disabled at 150 GeV. All BLM abort capabilities are disabled during acceleration and subsequent manipulations before collisions. On June 15, 2009, three individual loss monitors were configured and enabled at 980 GeV during a store for the Slow integration mode only to provide BLM abort capability. The loss monitors that were chosen are located directly downstream of collimators that form a limiting aperture during the store. Under the original BLM system, all loss monitors were disabled from issuing an abort during a store due to the fact that it had only one threshold and one integration type. These limitations resulted in a number of false aborts terminating long stores. Those problems were addressed with the new system making it reasonable to assign an abort threshold for use during colliding beam operation.

Figure 10 is an example of a store that was aborted due to a BLM abort at Tevatron service building E1. A power supply for a low-beta quadruple had tripped off during a store causing losses at E11. Losses progressed for about 31 ms as seen by the turn-by-turn trace on the right side of figure 10. When the BLM Slow sum threshold of 15 Rad/s was finally reached, the command to fire the abort kickers was sent and the beam was dumped. The significance of the E11 BLM abort is that a magnet quench did not accompany the abort. Typically, quenches occur in these scenarios until the correct BLM threshold can be determined to produce only a beam abort.

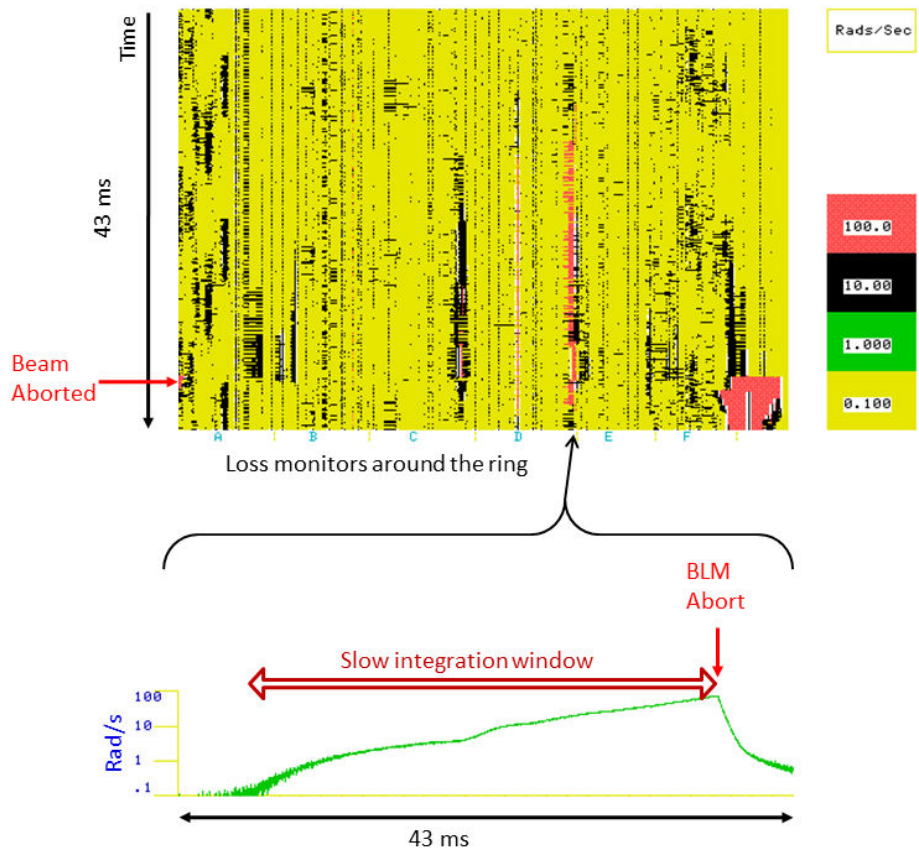

**Figure 10.** Example of beam abort from a loss monitor exceeding a threshold. The plot on the top is loss history (time increases going down) for each loss monitor around the ring (horizontal axis) during the ~43 ms prior to the beam abort. Red values are the largest and are greater than 100 Rad/s. The plot on the bottom shows the loss history of the loss monitor that pulled the abort. The Slow integration window is also shown. It is this Slow sum value that exceeded the 15 mRad/s limit and triggered the abort.

Another cause of beam aborts in the Tevatron is electrostatic separator sparks. The Tevatron stores protons and antiprotons in the same beam pipe. They are placed on helical orbits to avoid unwanted collisions thereby increasing the beam intensity lifetime. To separate the proton and antiproton orbits, there are 15 electrostatic separators operating at voltages of approximately 110 kV. These devices have a failure mode where the plates of the separator can generate a spark. The spark causes the separated orbits to come together causing large beam losses and magnet quenches. There have been roughly 32 quenches due to separator sparks during Collider Run II (2001 to 2011). These events are very fast, roughly a couple of nanoseconds to 10 microseconds, and were difficult to capture and diagnose with the previous BLM system. On November 17, 2010, there was a separator failure producing multiple sparks and multiple quenches, but this time BLM turn-by-turn data was captured by the new system. Figure 11 is a plot of the turn-by-turn losses at various Tevatron locations during this event. As seen on the red trace, LMD49 shows beam loss at turn 500 indicating the time of the first spark. The blue trace LME11 shows the second spark at turn 290. LME11 has its abort threshold enabled and the limit is exceeded resulting in a BLM generated abort at E1. The yellow trace is of LMA11 near the beam dump and shows the beam being aborted at turn 250. Unfortunately, the spark resulted in a quench at E1 due to the large amount of beam loss even though the BLM detected and pulled the abort prior to the quench. The BLM turn-by-turn system was able to record the timing of the sparks and quench events to provide an accurate diagnosis to the quench.

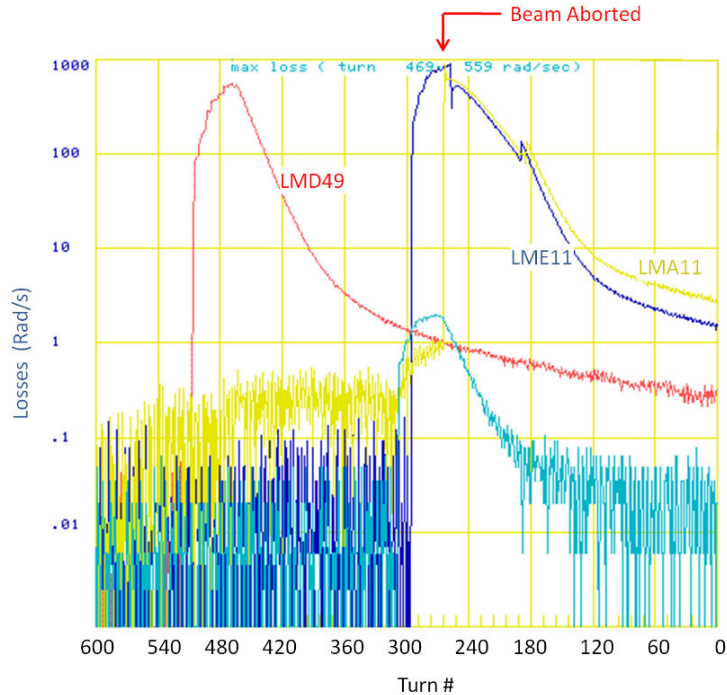

**Figure 11.** Loss structure from a Tevatron electrostatic separator spark. LMD49 shows the losses from the first spark and LME11 shows the losses from the second spark that triggered an abort. The losses were great enough to have generated a magnet quench despite the abort being pulled.

**5.2 Main Injector**

The upgraded BLM system in the Main Injector was installed in late 2006 and losses have been recorded on each Main Injector cycle since October 11, 2006. There are 250 loss monitors distributed around the 3319 m ring and mounted on the outer tunnel wall near the downstream end of the quadrupole magnet in each half cell. Several additional loss monitors are installed near the beam line at each transfer point.

Each BLM location provides an input to the Main Injector abort system. Loss sums with 40 ms sampling are compared with limits for each BLM and the beam is aborted if the limit is exceeded. Additionally, at the end of each cycle, the loss integral is compared to its limit and, if over threshold, beam of the next cycle is inhibited until the system is reset.

Beam loss plotting capability is available from each BLM system with a resolution up to 720 Hz. Using this feature, one can examine any loss feature in detail to correlate with other accelerator or beam time structure (figure 12). In addition to routine plotting for operations, special purpose data capture software can employ the same BLM electronics output to store time sequence data for detailed analysis on many channels.

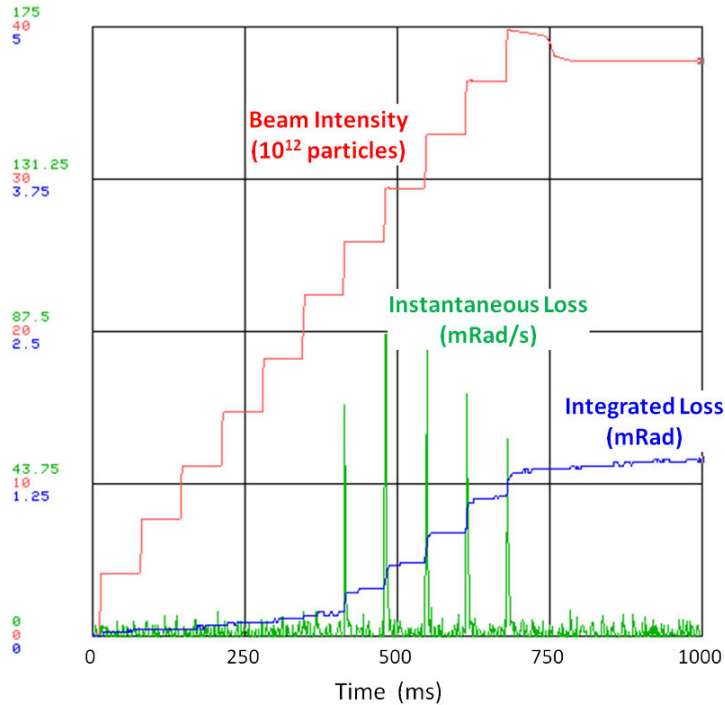

**Figure 12.** Loss structure in the Main Injector for normal neutrino and antiproton production cycles near the injection point. The beam in intensity in the Main Injector takes a step with each injection from the Booster. More injection losses are observed in the second half of the injection cycle due to slip stacking of the beam [14].

During routine operation, such as for antiproton and neutrino production cycles, sampling for the BLM system is set such that the integral losses for the cycle are stored at eight sample times. These values are used to produce a "comfort display", which shows the integral loss at the end of the cycle overlaid with the losses at the sample time corresponding to 1.5% acceleration (uncaptured beam loss) and also with the losses after injection [15] (figure 13). Using this logarithmic scale display, the Main Injector and operations staff are able to identify many unnecessary losses, including small ones, and tune orbits, acceleration parameters or other machine features to eliminate the loss.

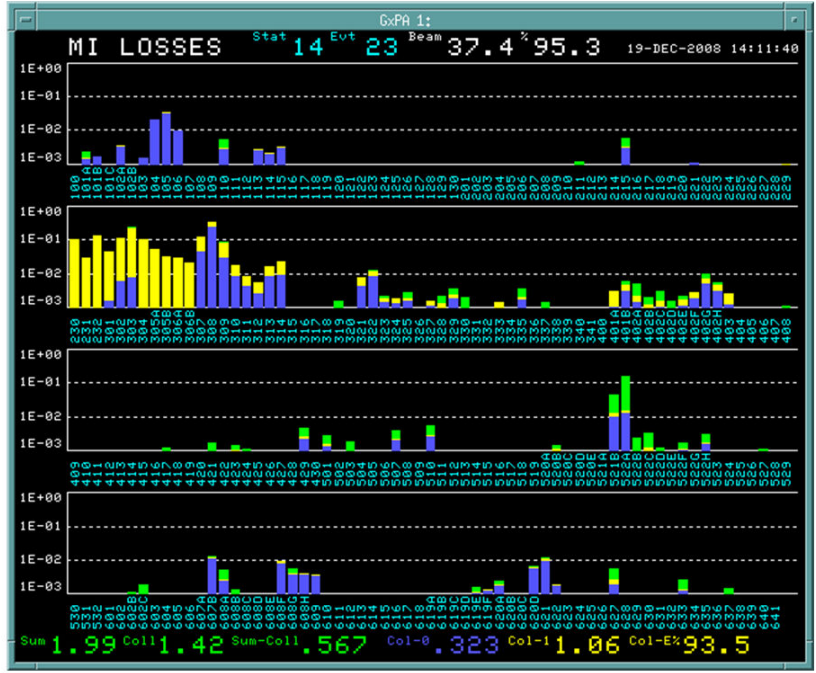

**Figure 13.** Main Injector comfort display showing, on a logarithmic scale, losses for all loss monitors at various times throughout a cycle. Losses integrated to the end of cycle are shown in green, losses after 1.5% acceleration in yellow and injection losses in blue. This display facilitates quickly comprehending the loss structure both in time and space.

The Main Injector slip stacking operation [14] produces losses due to beam which is not captured by the acceleration rf system. A collimation system to localize these losses was installed in 2007 [16,17]. In order to tune beam orbits and collimator positions and to measure the collimator effectiveness, additional features of the BLM system were employed. Sample times for the integral loss were set before and after each beam injection, after the uncaptured beam was lost, and at the end of the cycle. Single-cycle or multi-cycle averages of these losses were recorded for later analysis [15]. Alternatively, for studies, samples were obtained frequently during the uncaptured beam loss so that different loss mechanisms could be separated based on their time structure.

The losses recorded for the comfort display are also manipulated to provide summed values in the collimator region and for the ring. These values and the sum for each cycle for each BLM are recorded in the accelerator data logger system. Loss patterns and operational problems can be examined using this data. In addition, activation trends have been monitored using tunnel residual radiation measurements [18]. Residual radiation history has been successfully correlated with half-life weighted BLM history data to allow prediction of future activation results [19].

Using the data logged values, the effects of mismeasurements of the pedestal have been examined by selecting beam pulses with no accelerated beam and reading the integrated loss per cycle (in mRad). They have been studied in various 3 month time periods, using such pulses in the first 10,000 Main Injector cycles for each period. Using data for April 2010, a mean, $\mu_i$, and standard deviation, $\sigma_i$, of the observed pedestal is obtained for each BLM channel $i$. The distribution of the $\sigma_i$ has a mean value of $M_\sigma = 0.77$ mRad with a standard deviation of $S_\sigma = 0.12$ mRad, with only three channels having a $\sigma_i$ that is more than ($3S_\sigma$) away from $M_\sigma$. The mean and standard deviation of the $\mu_i$ are calculated to be $M_\mu = -0.78$ mRad and $S_\mu = 0.52$ mRad after iteratively applying $3S_\mu$ selection cuts to the $\mu_i$ to identify well behaved channels. About 16% (41 channels) of the $\mu_i$ are more than $3S_\mu$ above $M_\mu$. These channels with positive pedestal values will integrate to positive signals at cycle end and show a false loss on the comfort display. Only four channels (1.6%) have negative pedestals more than $3S_\mu$ below $M_\mu$ which will hide actual losses. The distribution of $\mu_i$ and $\sigma_i$ show little correlation although the two largest $\sigma_i$ also have more negative $\mu_i$. An examination of various 3-month periods finds that changes in the signal from drifting pedestal errors are in a range up to 5 mRad. Losses of 2 mRad will typically result in residual radiation in nearby accelerator components of 10 mRad/hr. This performance is adequate for control of residual radiation without needing to adjust for the pedestal errors of individual channels.

To achieve the 400 kW beam power capability required at present from the Main Injector, slip stacking injection, large aperture quadrupoles for transfer regions, beam dampers to control instabilities, collimators to localize losses, and fuzzers (anti-dampers) to remove beam from beam gaps required for kickers have been implemented. BLM systems have allowed these enhancements to be commissioned and tuned such that twice the beam power is available with residual radiation levels at many critical points being one tenth of their historical peak values.

## 6. Conclusion

The BLM upgrade is a significant improvement over the long serving previous system. The upgrade provides improved machine protection by implementing a usable abort on losses capability. The upgrade also provides better post abort diagnostics through its turn-by-turn buffer and the three buffers of integrated loss types. The upgrade has proven to be robust resulting in less downtime and maintenance requirements.

## Acknowledgments

The authors thank their colleagues throughout the Accelerator and Particle Physics Divisions for their assistance and co-operation in the production, testing, and installation of the upgraded beam-loss monitor systems. Their vigilance and responsiveness were essential to getting these systems up and running quickly and reliably.